\documentclass[proceedings, preprint]{rmaa}

\usepackage{paralist}

\usepackage{psfrag,color}

\SetYear{2008}
\SetConfTitle{Magnetic Fields in the Universe II}

\title{Polarization of FIR emission from T Tauri Disks} 

\author{
  Jungyeon Cho\altaffilmark{1}
  and A. Lazarian\altaffilmark{2}}

\altaffiltext{1}{Dept. of Astronomy and Space Science,
Chungnam National Univ., Korea (jcho@cnu.ac.kr).}

\altaffiltext{2}{Astronomy Dept., Univ. of Wisconsin,
  Madison, WI53706, USA (lazarian@astro.wisc.edu).}

\shortauthor{Cho \& Lazarian}
\shorttitle{Polarized emission from T Tauri Stars}

\listofauthors{Jungyeon Cho \& A. Lazarian}
\indexauthor{Cho, Jungyeon}
\indexauthor{Lazarian, A.}

\abstract{
Recent observation of 850$\mu$m sub-mm polarization 
{}from T Tauri disks opens up the possibility of studying
magnetic field structure within protostellar disks.
The degree of polarization is around 3 \% and
the direction of polarization is perpendicular to
the disk.
Since  thermal emission from dust grains dominates the spectral energy distribution at the sub-mm/FIR regime,
dust grains are thought to be the cause of the polarization.
We discuss grain alignment by radiation and we explore
the efficiency of dust alignment
in T Tauri disks.
Calculations show that dust grains located far away from the Central
proto-star are more efficiently aligned.
In the presence of a regular magnetic field, the aligned grains
produce polarized emission in sub-mm/FIR wavelengths.
The direction of polarization is perpendicular to the local
magnetic field direction.
When we use a recent T Tauri disk model and 
take a Mathis-Rumpl-Nordsieck-type distribution 
with maximum grain size of 500-1000 $\mu$m, 
the degree of polarization is
 around 2-3 \% level at wavelengths larger than $\sim$100$\mu$m.
Our study indicates 
that multifrequency infrared polarimetric studies of protostellar disks
can provide good insights into the details of their magnetic structure.
We also provide predictions for polarized
emission for disks viewed at different wavelengths and viewing angles. 
}

\addkeyword{polarization}
\addkeyword{accretion disk}
\addkeyword{dust, extinction}
\addkeyword{Stars: Pre-main sequence}

\begin{document}
\maketitle

\section{Introduction}
Recently, Tamura et al. (1999) first detected
polarized emission from T Tauri stars, low mass 
protostars. 
They interpreted the polarization (at $\sim 3$ \% level) 
in terms of thermal emission
from aligned dust grains.
Magnetic field is an essential component for grain alignment. 
If grains are aligned
with their long axes perpendicular to magnetic field, the
resulting grain emission has polarization directed perpendicular to the
magnetic field. 

The notion that the grains can be aligned in respect to magnetic field
can be traced back to the discovery of star-light polarization 
by Hall (1949) and Hiltner (1949), that arises from interstellar grains.
Historically the theory of the grain alignment was developing mostly to explain the interstellar
polarization, but grain alignment is a much wider spread phenomenon (see
Lazarian 2007 for a review). Among the alignment mechanisms the one 
related
to radiative torques (RTs) looks the most promising. We invoke it for our
calculations below.

The RTs make use of interaction of radiation with a grain 
to spin the grain up.
The RT alignment was first discussed by
Dolginov (1972) and Dolginov \& Mytrophanov (1976).
However, quantitative studies were done only in 1990's.
In their pioneering work, Draine \& Weingartner 
(1996, 1997) demonstrated the efficiency of the
RT alignment for a few arbitrary chosen
irregular grains using numerical simulations. This work identified RTs as
potentially the major agent for interstellar grain alignment.
A successful analytical model of RTs was suggested by Lazarian \& Hoang (2007).
 Cho \& Lazarian (2005) 
demonstrated the rapid increase of radiative torque efficiency and
showed that radiative alignment can
naturally explain decrease of the degree of polarization
near the centers of pre-stellar cores. Large grains are known to be present
in protostellar disk environments
 and this makes the RT alignment promising.

Roughly speaking, the efficiency of grain alignment by RTs depends on two factors - 
the intensity of radiation and the gaseous drag.
The latter depends on gas pressure.
Therefore, the ideal condition for grain alignment by RTs is strong radiation and 
low gas pressure.

In order to calculate efficiency of grain alignment, we need to know
radiation intensity, gas density, and temperature in T Tauri disks.
Recently proposed hydrostatic, radiative equilibrium passive
disk model
(Chiang \& Goldreich 1997; Chiang et al. 2001, hereafter C01)
fits observed SED from T Tauri stars very well
and seems to be one of the most promising models.
Here, passive disk means that active accretion effect, which might be 
very important in the immediate vicinity of the central star, is not 
included in the model.
In this paper we adopt the model in C01.

In this paper, we briefly discuss polarized FIR emission arising from 
aligned dust grains by radiative torque in T Tauri disks.
Detailed calculations and discussions can be found in 
Cho \& Lazarian (2007).
In \S II, we discuss grain alignment in T Tauri disks.
In \S III, we give theoretical estimates for degree of polarization.
In \S IV, we discuss observational implications.
We give summary in \S V.

\section{The disk model used for this study}
\begin{figure}[h!t]
\includegraphics[width=.38\textwidth]{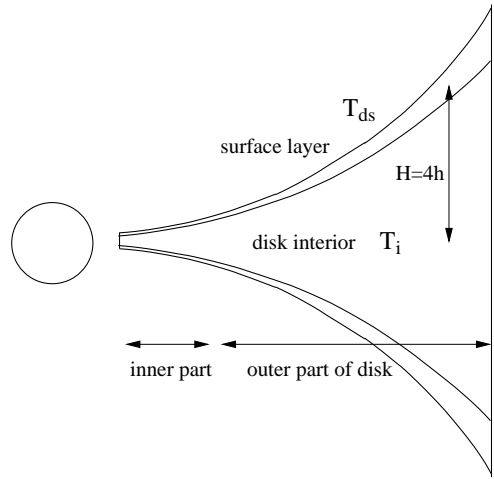}
\caption{ A schematic view of the disk model (see C01).
The surface layer is hotter and heated by the star light.
The disk interior is heated by re-processed light from the
surface layers. We assume that the disk height, $H$, is 4 time the
disk scale height, $h$. From Cho \& Lazarian (2007).
\label{fig:model}
}
\end{figure}

 We assume that magnetic field is regular and toroidal 
(i.e. {\it azimuthal}).
We use a T Tauri disk model in C01.
Figure \ref{fig:model} schematically  shows the model.
The disk is in hydrostatic and radiative equilibrium and shows flaring.
They considered a two-layered disk model. 
Dust grains in the surface layer
are heated directly by the radiation from the central star
and emit their heat more or less isotropically.
Half of the dust thermal emission immediately escapes and the other half
enters into disk interior and heats dusts and gas there.
They
assume that the disk interior is isothermal.

In our calculations, we use a grain
model similar to that in C01.
We use an MRN-type power-law distribution of grain radii $a$ between
$a_{min}$ (=0.01 $\mu m$ for both disk interior and surface layer)
 and $a_{max}$ (=1000 $\mu m$ for disk interior and $=1\mu m$
for disk surface layer)
with a power index of -3.5: $dN \propto a^{-3.5} da$.
As in C01 we assume that grain composition varies 
with distance from the central star
in both disk interior and surface layer.
We assume that grains in the surface layer 
are made of silicate only when the distance 
$r$ is less than 6 AU, and silicate covered with water ice 
when $r>6$AU.
We do not use iron grains for the immediate vicinity of
the star.
We assume that grains in the disk interior are made of 
silicate when $r<0.8$AU and ice-silicate for $r>0.8$AU.
 The fractional thickness of the water ice mantle,
$\Delta a/a$, is
set to 0.4 for both disk surface and disk interior.
Unlike C01, we use the refractive index of astronomical silicate 
(Draine \& Lee 1984; Draine 1985;
Loar \& Draine 1993; see also Weingartner \& Draine 2001).
We take optical constants of pure water ice from a NASA web site
(ftp://climate1.gsfc.nasa.gov/wiscombe).

The column density of the disk is $\Sigma_0 r_{AU}^{-3/2}$ with
$\Sigma_0=1000 g/cm^2$. 
Here $r_{AU}$ is distance measured in AU.
The disk is geometrically flared and 
the height of the disk surface is set to 4 times the disk scale height $h$.
The disk inner radius is $2R_*$ and the outer radius is $100$AU.
The central star has radius of $R_* = 2.5 R_{Sun}$ and temperature
of $T_* = 4000$K.
Temperature profile, flaring of disk, and
other details of the disk model
are described in C01.

\section{Grain alignment by radiation}

\subsection{Polarized FIR emission from aligned grains}

\begin{figure*}[h!t]
\includegraphics*[width=.45\textwidth,angle=-90]{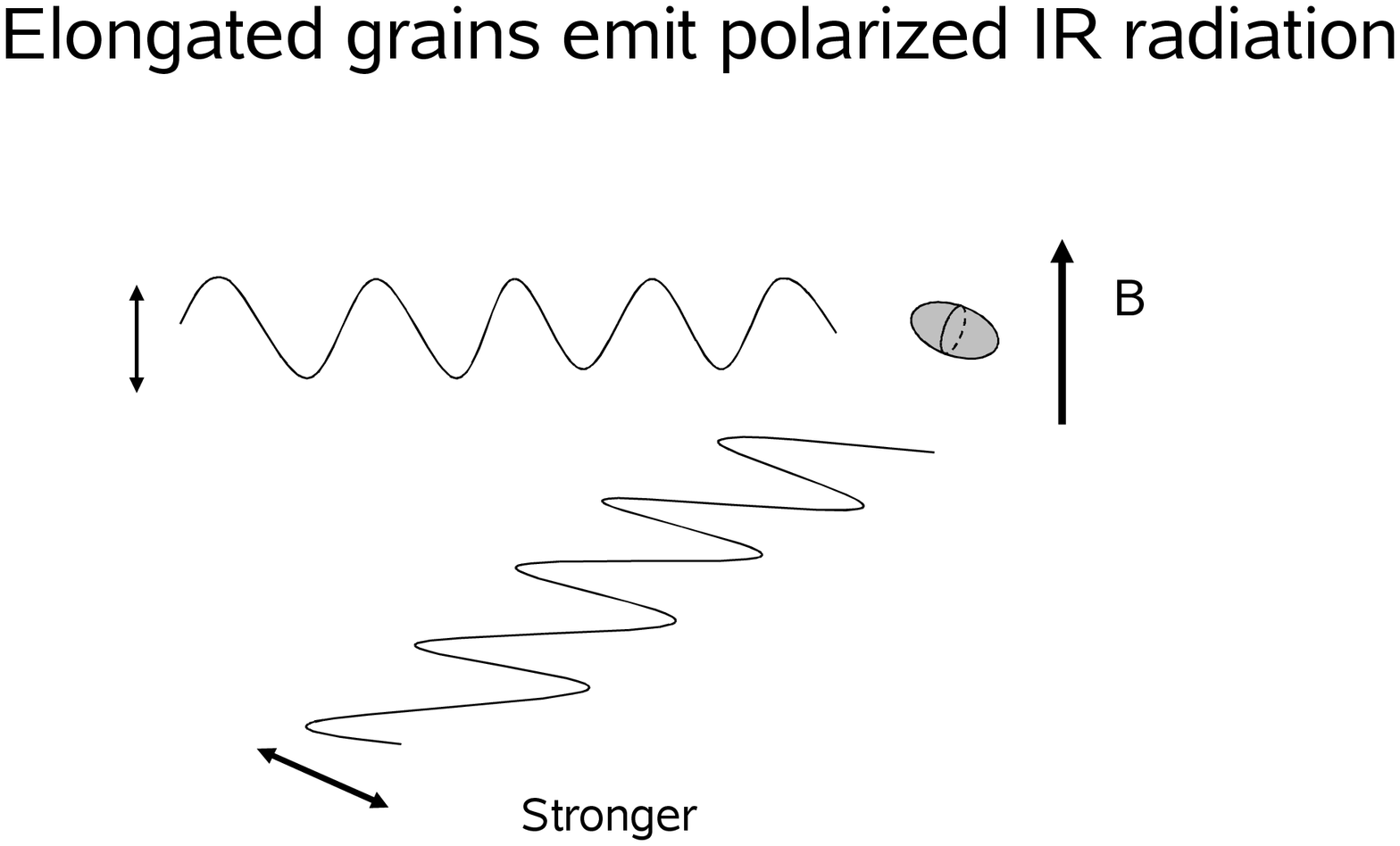}
\hspace{-3cm}
\includegraphics*[width=.45\textwidth,angle=-90]{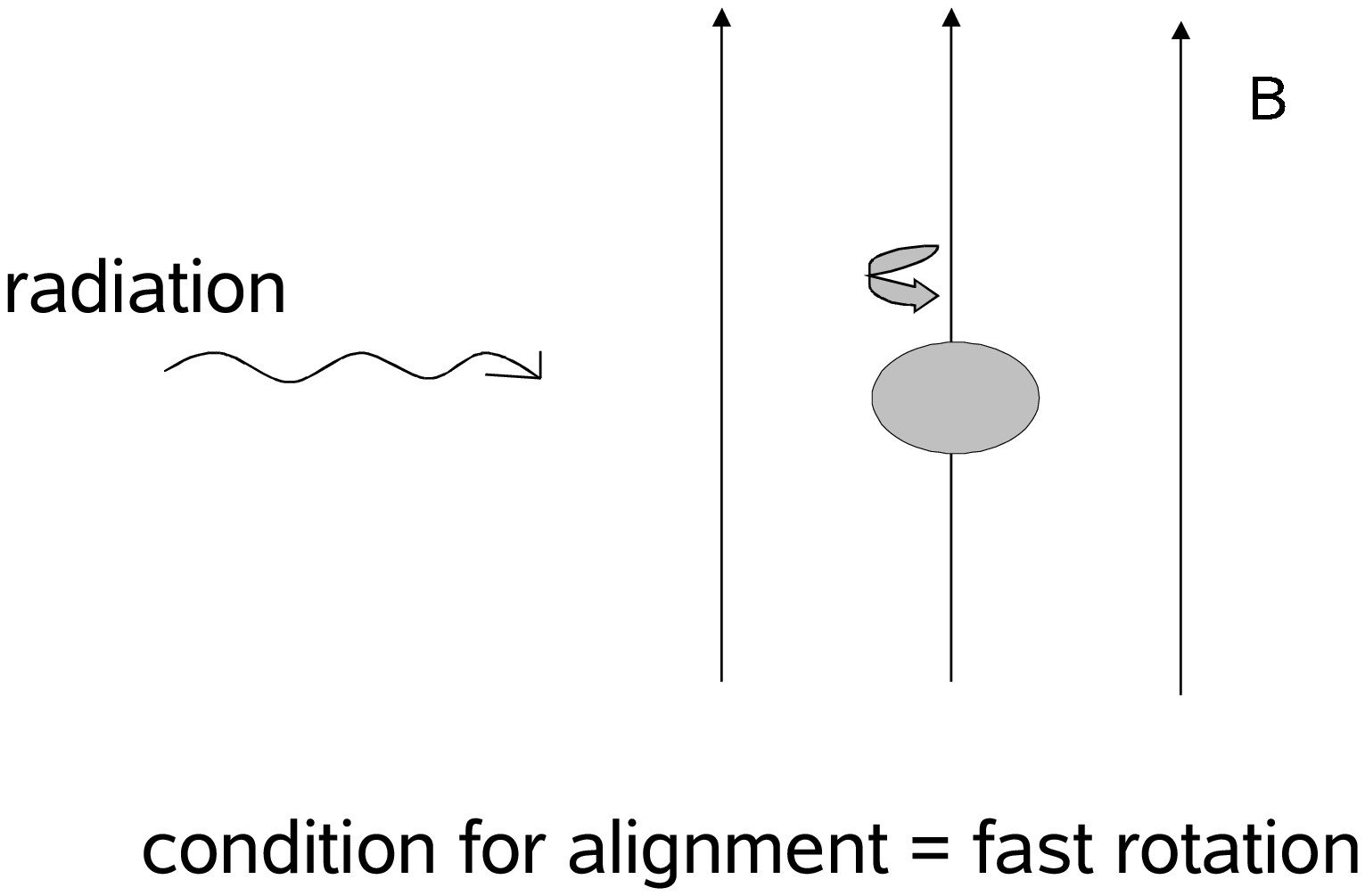}
\caption{Grain alignment and polarized emission. 
{\it Left}: Emission from an elongated grain. When electric field is parallel
to the grain's long axis, the radiation is stronger.
{\it Right}: Conditions for grain alignment. Roughly speaking,
low gas density and strong rotation
are favorable conditions for grain alignment.}
\label{fig:emission}
\end{figure*}

Theories predict that grain alignment happens in such a manner that
grain's long axis is perpendicular to local magnetic field direction
(see Figure \ref{fig:emission}).
Grains are usually cold and emit infrared (IR) radiation.
When an elongated grain emit IR radiation, the one with 
electric vector parallel to the grain's long axis is stronger 
(left panel of Figure \ref{fig:emission}).
Therefore the direction of polarization is parallel to the grain's
long axis, or perpendicular to the magnetic field.

Then what is the condition of grain alignment?
Roughly speaking, fast rotation is a necessary condition for
grain alignment. 
Then what is the condition for fast rotation?
Of course, when the intensity of radiation is strong, 
grains can rotate faster.
Gases density is also an important factor.
Since gaseous drag slows down grain's rotation, low gas density
(more precisely, gas pressure) is a favorable condition for 
fast rotation.
(Note however that this is just a possible
parameterzation for RTs. See Lazarian \& Hoang 2007 and
the last paragraph of this section for details.)

\subsection{Radiative torque for large grains}

\begin{figure}[h!t]
\includegraphics[width=.48\textwidth]{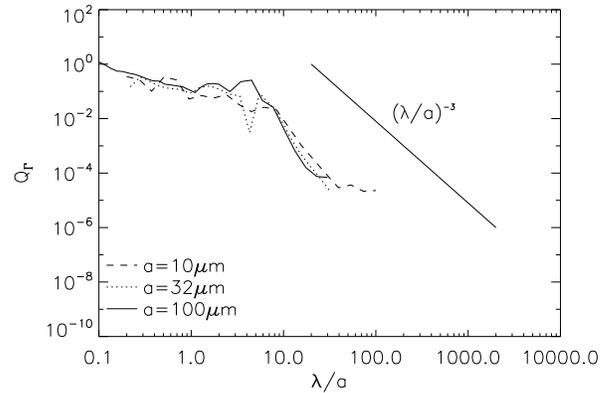}
\caption{
Behavior of Torque.
Torque is $\sim O(1)$ when $\lambda \sim a$, 
where $a$ is the grain size.
Roughly speaking, torque $\propto (\lambda/a)^{-3}$.
The results for large grains.
    Data from Lazarian \& Hoang (2007).
\label{fig:tq}
}
\end{figure}

For most of the ISM problems,
dust grains are usually smaller than the wavelengths of interest.
However, this is no longer true in T Tauri disks because
we are dealing with grains as large as $\sim 1000 \mu m$.
To understand grain alignment in T Tauri disks we need to know
radiative torque for large grains.
Motivated by Figure \ref{fig:tq}, we assume
that the radiative torque 
\begin{equation}
 \Gamma_{rad} = \pi a_{}^2 u_{rad}\frac{\lambda}{2\pi} Q_\Gamma
\end{equation}
with
\begin{eqnarray}
 Q_\Gamma = \left\{ \begin{array}{ll}
                    \sim O(1) & \mbox{~~~ if $\lambda \sim a$} \\
                    \sim (\lambda/a)^{-3} & \mbox{~~~if $\lambda > a$,}
                    \end{array}
             \right.
\label{eq:tq}
\end{eqnarray}
where $a$ is the grain size, $u_{rad}$ the energy density
of the incident radiation,
 and $\lambda$ the wavelength of the incident
radiation. Note that
$Q_\Gamma \sim O(1)$ when $\lambda \sim a$.

\subsection{Rotation rate of dust grains by radiative torque}
We assume that to align grains RTs should spin grains suprathermally.
Detailed theory of grain alignment can be more complicated
(see recent study by Lazarian \& Hoang 2007, 2008; Hoang \& Lazarian 2007).

After some modifications, equation (67) in Draine \& Weingartner (1996)
 reads
\begin{eqnarray}
\left( \frac{ \omega_{rad} }{ \omega_T } \right)^2
 = 4.72 \times 10^9 \frac{ \alpha_1 }{ \delta^2 } \rho_3 a_{-5} 
  \left( \frac{ u_{rad} }{ n_H kT } \right)^2
  \left( \frac{ \lambda }{ \mu m } \right)^2  \\ \nonumber
   [ Q_{\Gamma} ]^2 
  \left( \frac{ \tau_{drag} }{ \tau_{drag,gas} } \right)^2,
\label{eq:ww}
\end{eqnarray}
where $Q_{\Gamma} = \bf{Q}_{\Gamma} \cdot \hat{\bf{a}}_1$ and
$\hat{\bf{a}}_1$ is the principal axis with largest moment of
inertia, $n_H$ is the hydrogen number density, 
$u_{rad}$ is the energy density of the radiation field, 
$\delta \approx 2$, $\alpha_1 \approx 1.745$,
$\rho_3=rho/3g cm^{-3}$, $a_5=a/10^{-5}cm$,
and $\omega_T$ is the thermal angular frequency, which is the rate
at which the rotational kinetic energy of a grain is equal to $kT/2$.
The timescales 
$\tau_{drag,gas}$ and $\tau_{drag,em}$ are the damping time for gas drag and for electromagnetic emission, respectively, and
they satisfy the relation
$\tau_{drag}^{-1}=\tau_{drag,em}^{-1}+\tau_{drag,gas}^{-1}$  (see
Draine \& Weingartner 1996 for details).
As we discussed in the previous subsection, $Q_{\Gamma}$ is of
order of unity when $\lambda \sim a$ and declines as
$(\lambda/a)$ increases. From this observation, we can write
\begin{eqnarray}
\left( \frac{ \omega_{rad} }{ \omega_T } \right)^2  
\approx 
\left( \frac{ \omega_{rad} }{ \omega_T } \right)_{\lambda \sim a}^2
\left( \frac{ Q_{\Gamma, \lambda \sim a} }
            { Q_{\Gamma, \lambda}        } \right)^2  \nonumber \\ 
\approx 
\left( \frac{ \omega_{rad} }{ \omega_T } \right)_{\lambda \sim a}^2
\left( \frac{ \lambda  }{a} \right)^{-6}
\label{eq:tqlambda}
\end{eqnarray}
for $\lambda > a$, where
\begin{eqnarray}
\left( \frac{ \omega_{rad} }{ \omega_T } \right)_{\lambda \sim a}^2
 \approx 4.72 \times 10^9 \frac{ \alpha_1 }{ \delta^2 } \rho_3 a_{-5} 
  \left( \frac{ u_{rad} }{ n_H kT } \right)^2  \nonumber \\ 
  \left( \frac{ \lambda }{ \mu m } \right)^2  
  \left( \frac{ \tau_{drag} }{ \tau_{drag,gas} } \right)^2,
  \label{eq:tqla}
\end{eqnarray}

The limitation of this approach is that the amplitude values of the RTs are used 
to parameterize the alignment.
In fact, Lazarian \& Hoang (2007) showed that the RTs amplitude may change
substantially with the angle between the radiation direction and magnetic field.

\begin{figure}[h!t]
\includegraphics[width=.48\textwidth]{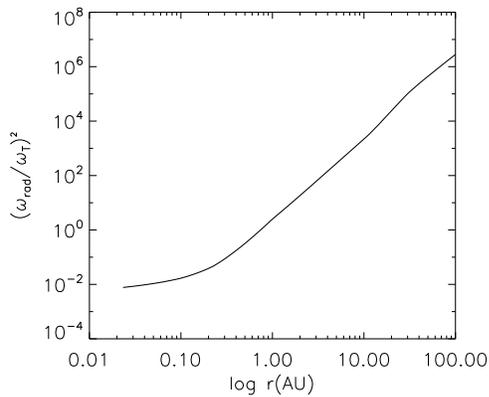}
\caption{Grain alignment in surface layer. 
The ratio 
$(\omega_{rad}/\omega_T)_{\lambda \sim a}^2$
exceeds 10 when $r \geq 1$AU, which means that
some grains in the surface layer are aligned when  $r \geq 1$AU.
Results are for $a=1\mu$m grains. {}From Cho \& Lazarian (2007).
\label{fig:align_sur}
}
\end{figure}
\begin{figure}[h!t]
\hspace{-1.5cm}
\includegraphics[width=.45\textwidth,angle=-90]{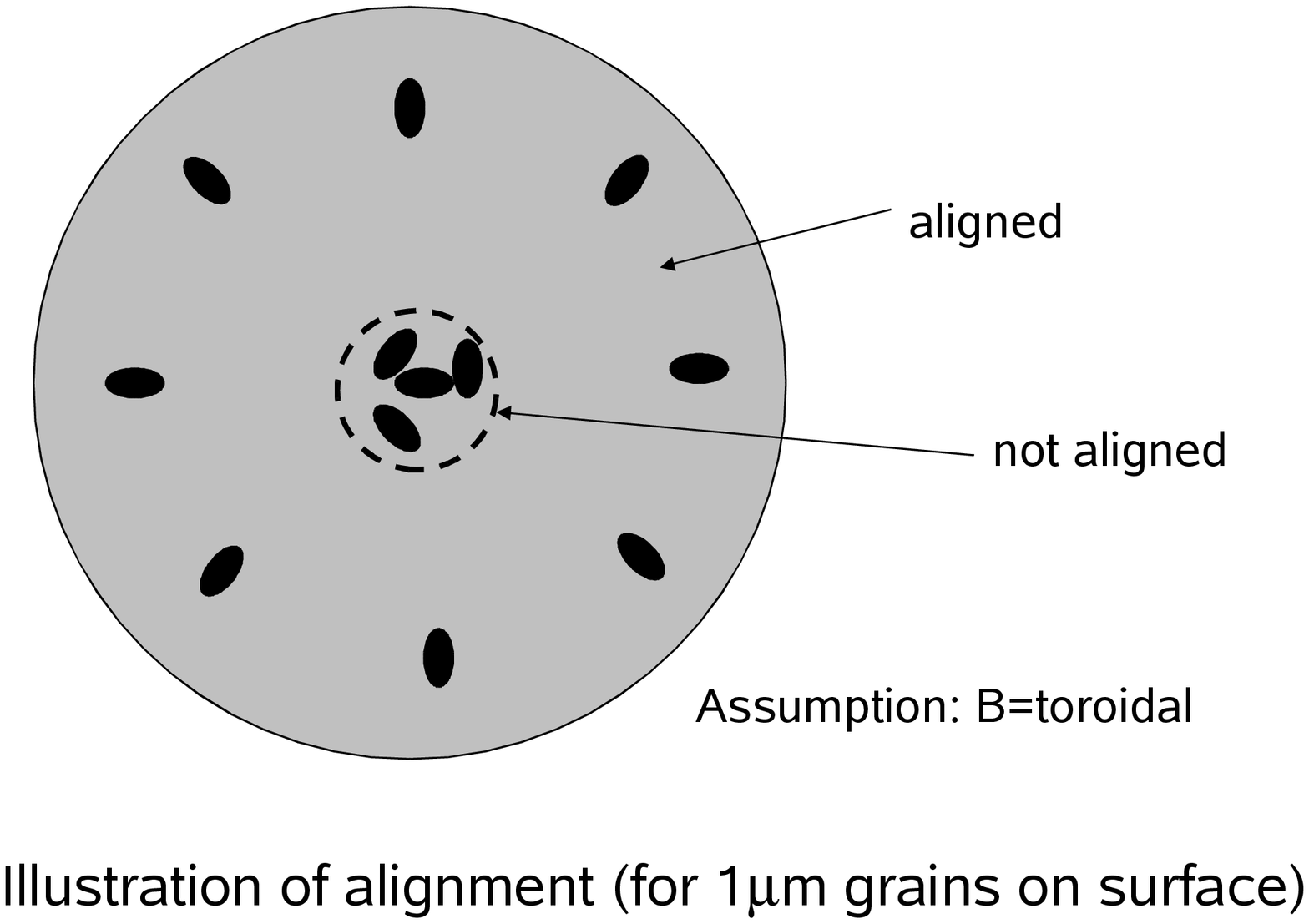}
\caption{Illustration of grain alignment on disk surface.
 Near the star, grains are not aligned due to high density.}
\label{fig:illustration}
\end{figure}

\section{Grain alignment in disks}
We use Eq.~(\ref{eq:tq}), instead of the DDSCAT software package, 
to obtain radiative torque ($Q_{\Gamma}$) on grain particles in the T Tauri disks.
We take a conservative value of $Q_{\Gamma}$ at $\lambda \sim a$:
 $Q_{\Gamma}\sim 0.1$ at $\lambda \sim a$.
Apart from $Q_{\Gamma}$, we also need to know $u_{rad}$ and 
$n_H$ to get the $(\omega_{rad}/\omega_T)_{\lambda \sim a}^2$ ratio
(see Eq.~(\ref{eq:tqla})). 
We directly calculate $u_{rad}$ and 
$n_H$ using the disk model in C01.
We assume that $\tau_{drag}\sim \tau_{drag,gas}$.

We assume that 
the RT alignment is perfect when
the ratio 
$(\omega_{rad}/\omega_T)_{\lambda \sim a}^2$ exceeds 10\footnote{
The perfect alignment is true for grains having superparamagnetic inclusions
(Lazarian \& Hoang 2008).
For ordinary paramagnetic grains the degree of alignment
can vary.}.
Calculations (see Cho \& Lazarian 2007) show that 
grains near the central star cannot be aligned due to
high gas density near the star.
Indeed Figure \ref{fig:align_sur} shows the ratio 
$(\omega_{rad}/\omega_T)_{\lambda \sim a}^2$
exceeds 10 when the distance from the central star, $r$, 
is large, which means that
grains in the surface layer are aligned when  $r$ is large. 
We expect that polarized emission from the surface layer
is originated from outer part of the disk.
Similar results hold true for disk interior.
Calculations show that, at large $r$, large grains 
are aligned even deep inside the interior.
On the other hand, at small $r$, only grains near the
disk surface are aligned 
(see an illustration in Figure \ref{fig:illustration}).

\begin{figure}[h!t]
\includegraphics[width=.50\textwidth]{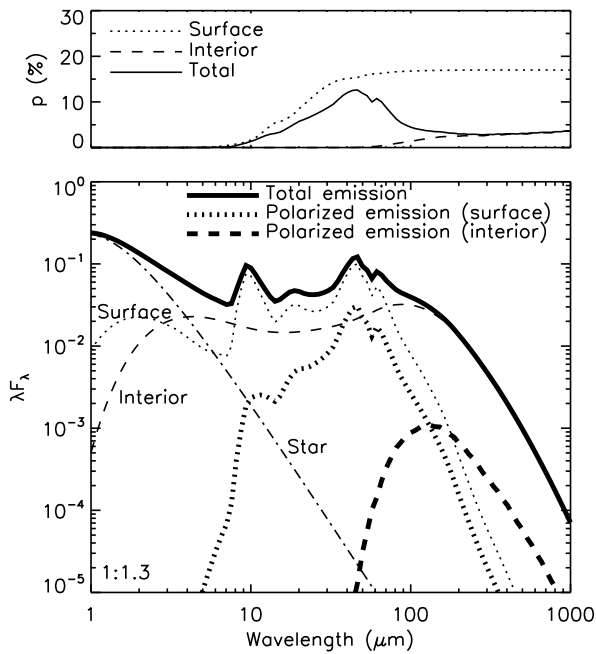}
\caption{Spectral energy distribution.
The vertical axis (i.e. $\lambda F_{\lambda}$)
is in arbitrary unit.
Thick solid line: total (i.e.~interior + surface) emission from disk.
Thin dotted line: total emission from disk surface.
Thick dotted line: polarized emission from disk surface.
Thin dashed line: total emission from disk interior.
Thick dashed line: polarized emission from disk interior.
Note that,  in these calculations of polarized emission, 
we ignored the direction
of polarization vectors and we only take the absolute value of them.
  Results for oblate spheroid grains with axis ratio of 1.3:1.
{}From Cho \& Lazarian (2007).
\label{fig:sed16}
}
\end{figure}
\section{Predictions for Degree of Polarization}

\subsection{Estimates for Spectral Energy Distribution}
In this subsection, we calculate the degree of polarization of
emitted infrared radiation from a disk with structure and 
parameters described in
C01.
In this subsection, we assume that the disk in face-on.
The degree of polarization will be zero for a face-on disk when
magnetic field is perfectly azimuthal 
and the disk is cylindrically symmetric.
In this section, {\it we are concerned only with the
 absolute magnitude of the polarization}.

Figure \ref{fig:sed16}  
shows the results for 1.3:1 oblate spheroid.
The degree of polarization can be as large as $\sim$5\% in FIR/sub-millimeter
wavelengths and $\sim$ 10\% in mid-IR regimes.
The polarized emission at FIR is dominated by the disk interior and
that at mid-IR is dominated by the disk surface layer.
Note again that, in these calculations, we ignored the direction
of polarization and we only take the absolute value of it.
 Note that, 
since the degree of polarization of emission from the disk surface layer is
very sensitive to the maximum grain size in the surface layer,
the results for $\lambda < 100 \mu m$ should be very sensitive to
the maximum grain size in the surface layer.

\subsection{Radial energy distribution}

\begin{figure*}[h!t]
\includegraphics[width=.95\textwidth]{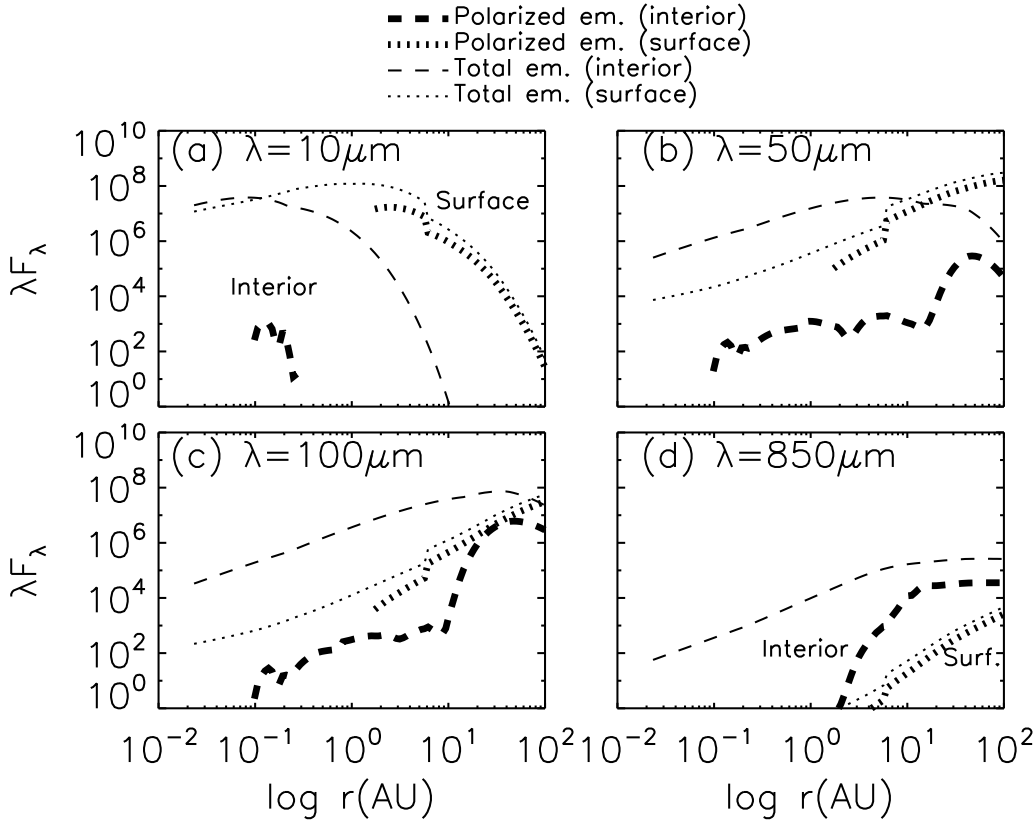}
\caption{Radial energy distribution.
(a) $\lambda=10 \mu m$. Inner part of the disk
   emits substantial amount of radiation. 
   But it emits negligible amount of polarized radiation.  
   Note that, when $r<1$AU, grains in the surface layer
   are not aligned and only negligible fraction of
   grains are aligned in the interior
   (see Figure   
     \ref{fig:align_sur} for disk surface).
(b) $\lambda=50 \mu m$.
(c) $\lambda=100 \mu m$.  
(d) $\lambda=850 \mu m$.
 The result for $\lambda=450 \mu m$ (not shown) is very similar to
 that for  $\lambda=850 \mu m$. {}From Cho \& Lazarian (2007).
\label{fig:red}
}
\end{figure*}

Figure \ref{fig:red} shows radial distribution of emitted radiation.
For $\lambda=850 \mu m$, both radiations from the disk interior and
the surface layer are dominated by the outer part of the disk.
But, for $\lambda=10 \mu m$, the inner part of the disk
contributes significant portion of total emission and polarized emission.

\subsection{Effects of disk inclination}

In this subsection we calculate actual degree of polarization
that we can observe.
Chiang \& Goldreich (1999) calculated spectral energy
distribution (SED)
from inclined disks.
We follow a similar method to calculate the
the SED of
polarized emission.

\begin{figure*}[h!t]
\includegraphics[width=.98\textwidth]{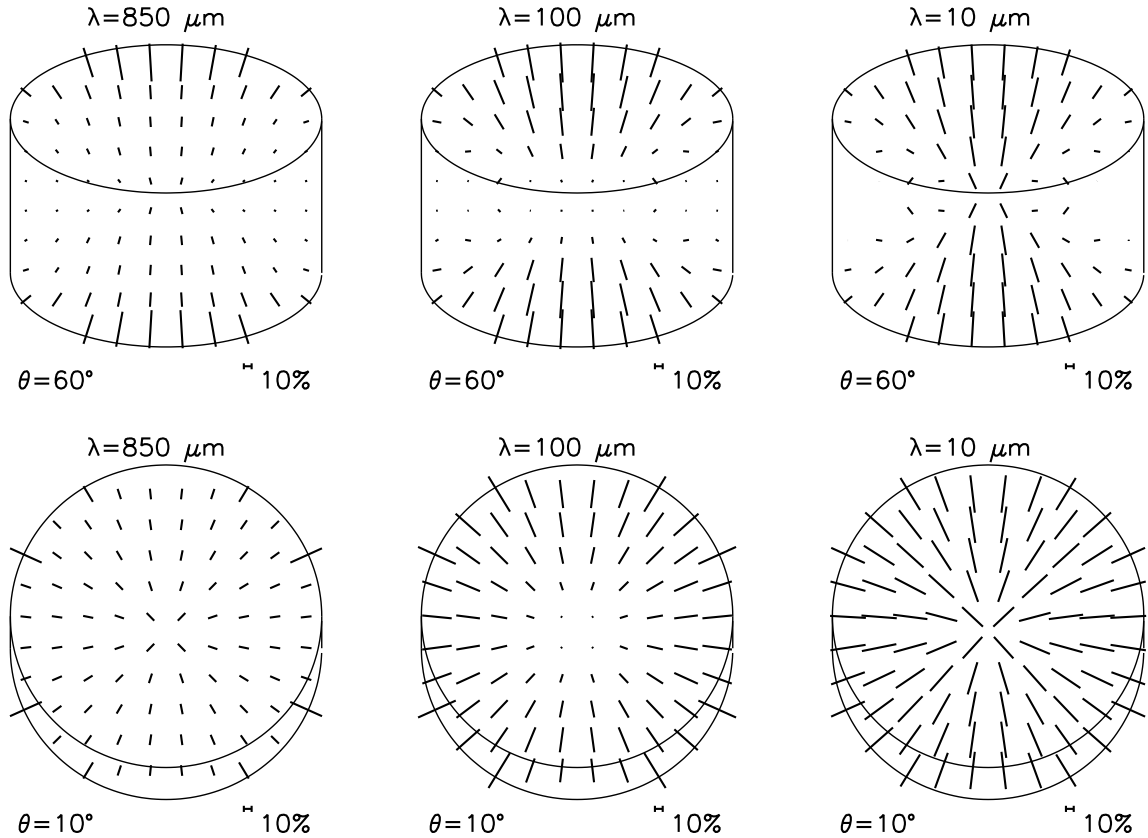}
\caption{
  Simulated observations. Degree of polarization is calculated for
the total radiation (i.e. interior + surface) from the disk.
 The disk inclination angle $\theta$ is the angle between
 disk symmetry axis and the line of sight.
{}From Cho \& Lazarian (2007).
\label{fig:visual}
}
\end{figure*}

Figure \ref{fig:visual}  
shows the effects of the disk inclination.
We calculate the polarized emission from the disk interior.
The viewing angle $\theta$ (=the angle of disk inclination) is the angle between the disk symmetry axis
and the line of sight.
We plot the direction of polarization for 3 different wavelengths and
2 different viewing angles.
The lines represent the direction of polarization.
Since we assume that magnetic field is azimuthal, the
direction of polarization is predominantly radial (see lower panels).

\section{Prospects}
Multifrequency observations of protostellar disks have become a booming field recently.
As grains at different optical depths have different temperatures,
multifrequency measurements reveal the structure of the disk.
They have advanced substantially our knowledge of the disks and allowed theoretical
expectations to be tested. 

Our study reveals that multifrequency polarimetry is very important for the protostellar
disks. The synthetic observations that we provide explicitly show that observations
at wavelength less than 100 $\mu$m mostly test magnetic fields of the skin layers, while
at longer wavelengths test magnetic fields of the bulk of the disk. Therefore polarimetry
can, for instance, test models of accretion, e.g.
 layered accretion (Gammie 1996).
Combining the far-infrared polarimetry with polarimetric measurements at
different frequencies may provide additional insight into the magnetic
properties of protostellar accretion disks. 

Most of the present day polarimetry will be done for
not resolved 
protostellar disks.
The size of the 
T Tauri disks is usually less than
$\sim 300$ AU (see, for example, C01). 
If we take the distance to proto-stars to be around $\geq 100 pc$,
then the angular sizes of the disks are usually smaller than
$6\arcsec$. 
The angular resolution of SCUBA polarimeter (SCUPOL) is around
$14\arcsec$ (Greaves et al. 2000) and that of SHARC II polarimeter
(SHARP; Novak et al. 2004) at $350 \mu m$ is around $9\arcsec$.
Therefore it is not easy to obtain plots like Figures \ref{fig:visual}.
The angular resolution of the intended SOFIA polarimeter is 
around $5\arcsec$ at $53 \mu m$, 
$9\arcsec$ at $88 \mu m$, and 
$22\arcsec$ at $215 \mu m$.
We see that the intended SOFIA polarimeter will be
 at the edge of resolving structure of close-by disks, while other instruments
will not resolve typical T Tauri disk. Therefore for most
of the near future observations
our predictions in Fig.~\ref{fig:sed16} and \ref{fig:visual} are most relevant.

\section{Summary}

Making use of the recent advances in grain alignment theory
we calculated grain alignment by RTs in a magnetized T Tauri disk.
Based on this, we calculated polarized emission from the disk.
Our results show that
\begin{itemize}

\item Polarization arising from aligned grains reveals magnetic fields
of the T Tauri disk. 

\item Disk interior dominates polarized emission in FIR/sub-millimeter
wavelengths. 

\item Disk surface layer dominates polarized emission in mid-IR wavelengths.
The degree of polarization is very sensitive to the
maximum size of grain in the disk surface layer.

\item Our study of the effect of the disk inclination predicts substantial
changes of the degree of polarization with the viewing angle.
The coming mid-IR/FIR polarimeters are very promising for studies of magnetic
fields in protostellar disks.

\item Polarization at different wavelengths reveals
      aligned grains at different optical depths, which allows one to
      tomography magnetic field structure.
 
\end{itemize}


\acknowledgments
Jungyeon Cho's work was supported by 
Korea Foundation for International Cooperation of Science \&
Technology (KICOS) through the Cavendish-KAIST Research Cooperation 
Center.
A. Lazarian acknowledges the support 
by the NSF grants AST 02 43156 and AST 0507164, as well as by the
NSF Center for Magnetic Self-Organization in Laboratory and Astrophysical 
Plasmas.


\end{document}